\documentclass{aa}

\makeatletter
\renewcommand*\aa@pageof{, page \thepage{} of \pageref*{LastPage}}
\makeatother

\usepackage{graphicx}
\usepackage[normalem]{ulem}
\usepackage{newtxtext}
\usepackage[smallerops,cmbraces]{newtxmath}
\usepackage[usenames,dvipsnames]{xcolor}
\definecolor{xlinkcolor}{cmyk}{1,1,0,0}
\usepackage[
  bookmarks=true,         
  pdfnewwindow=true,      
  colorlinks=true,        
  linkcolor=xlinkcolor,   
  citecolor=xlinkcolor,   
  filecolor=xlinkcolor,   
  urlcolor=xlinkcolor,    
  final=true,
]{hyperref}

\usepackage{bm} 
\usepackage{textgreek} 

\usepackage[capitalize]{cleveref} 
\crefname{section}{Sect.}{Sects.}
\creflabelformat{equation}{#2#1#3} 
\crefname{enumi}{item}{items} 

\usepackage{siunitx} 
\DeclareSIUnit[number-unit-product = ]\percent{\char`\%} 

\usepackage{csquotes} 

\usepackage{dblfloatfix}
\usepackage{adjustbox}
\usepackage{subcaption}
\definecolor{blackberry}{HTML}{8D1D75}

\newcommand*{\ngc}[1]{\object{NGC\,#1}}

\newcommand*{\abell}[1]{\object{Abell\,#1}}

\DeclareSIUnit\parsec{pc}

\DeclareSIUnit\kiloparsec{kpc}
\DeclareSIUnit\dex{dex}
\DeclareSIUnit\h{\mathnormal{h}}
\DeclareSIUnit\year{yr}
\DeclareSIUnit\years{yrs}
\DeclareSIUnit\arcsec{arcsec}
\DeclareSIUnit\arcmin{arcmin}
\DeclareSIUnit\Msun{M_\odot}
\DeclareSIUnit\Rsun{R_\odot}
\DeclareSIUnit\Lsun{L_\odot}
\DeclareSIUnit\Rvir{\mathnormal{R}_\mathrm{vir}}
\DeclareSIUnit\Rhalf{\mathnormal{R}_{1/2}}
\DeclareSIUnit\erg{erg}
\DeclareSIUnit\angstrom{\text{Å}}

\newcommand*{\Msun}{\ensuremath{\mathrm{M}_\odot}} 
\newcommand*{\Rsun}{\ensuremath{\mathrm{R}_\odot}} 
\newcommand*{\Lsun}{\ensuremath{\mathrm{L}_\odot}} 
\newcommand*{\Rvir}{\ensuremath{R_\mathrm{vir}}} 
\newcommand*{\Rhalf}{\ensuremath{R_{1/2}}} 
\newcommand{\e}[1]{\times 10^{#1}~}

\definecolor{lightblue}{rgb}{0.1,0.5,0.89}

\newcommand{\refadd}[1]{{\color{black}#1}}
\newcommand{\laned}[1]{{\color{black}#1}}

\begin{document}

\title{Merge and \laned{strip} -- \laned{dark matter-free dwarf galaxies} in \laned{clusters can be formed} by \laned{galaxy mergers}}
\titlerunning{Merge and Strip}

\author{
    Anna Ivleva\inst{\ref{inst:usm}} 
    \and
    Rhea-Silvia Remus\inst{\ref{inst:usm}}
    \and
    Lucas M.\ Valenzuela\inst{\ref{inst:usm}}
    \and
    Klaus Dolag\inst{\ref{inst:usm},\ref{inst:mpa}}
}
\authorrunning{A.\ Ivleva et al.}

\institute{
    Universitäts-Sternwarte, Fakultät für Physik, Ludwig-Maximilians-Universität München, Scheinerstr.\ 1, 81679 München, Germany\label{inst:usm}\\
    \email{ivleva@usm.lmu.de}
    \and
    Max-Planck-Institut für Astrophysik, Karl-Scharzschild-Str.\ 1, 85748 Garching, Germany\label{inst:mpa}\\
}

\date{Received XX Month, 20XX / Accepted XX Month, 20XX}

\abstract
{Recent observations of galaxy mergers inside galaxy cluster environments, such as \ngc{5291} in the vicinity of \abell{3574}, report high star formation rates in the ejected tidal tails, which point towards currently developing tidal dwarf galaxies. This prompts the intriguing question \refadd{whether} these newly formed stellar structures could get stripped from the galaxy potential by the cluster and thus populate it with dwarf galaxies.}
{\refadd{We verify whether environmental stripping of tidal dwarf galaxies from galaxy mergers inside galaxy cluster environments is a possible evolutionary channel to populate a galaxy cluster with low-mass and low surface brightness galaxies.}}
{\refadd{We \laned{performed} three high-resolution hydrodynamical simulations of mergers between spiral galaxies in a cluster environment, implementing a stellar mass ratio of 2:1 with $M_{200} = 9.5\e{11} \, \Msun$ for the more massive galaxy. Between the three different simulations, we \laned{varied} the initial orbit of the infalling galaxies with respect to the cluster center.}}
{We demonstrate that cluster environments are capable of stripping tidal dwarf galaxies from the host potential independently of the infall orbit of the merging galaxy pair, without instantly destroying the tidal dwarfs. Starting to evolve separately from their progenitor, these newly formed dwarf galaxies reach total masses of $M_\text{tot} \approx 10^{7-9} \, \Msun$ within the limits of our resolution. In the three tested orbit scenarios, we find \laned{three, seven, and eight} tidal dwarf galaxies per merger, respectively, which survive longer than \SI{1}{\giga\year} after the merger event. Exposed to ram pressure, these gas dominated dwarf galaxies exhibit high star formation rates while also losing gas to the environment. Experiencing a strong headwind due to their motion through the intracluster medium, they quickly lose momentum and start spiraling towards the cluster center, reaching distances on the order of \laned{\SI{1}{\mega\parsec}} from their progenitor. About \SI{4}{\giga\year} after the merger event, we still find \laned{three and four} intact dwarf galaxies in two of the tested scenarios, respectively. The other stripped tidal dwarf galaxies either evaporate in the hostile cluster environment due to their low initial mass, or are disrupted as soon as they reach the cluster center.}
{The dwarf production rate due to galaxy mergers is elevated when the interaction with a cluster environment is taken into account. Comparing their contribution to the observed galaxy mass function in clusters, our results indicate that \mbox{$\sim$30\%} of dwarf galaxies in clusters could have been formed by stripping from galaxy mergers.}


\maketitle
%

\section{Introduction}
\label{sec:introduction}
Observations and simulations suggest that dwarf-sized galaxies contain a significant amount of mass in the galaxy mass function of the local Universe \citep[e.g.,][]{sabatini+03, schaye+16}. Nevertheless, the variety of formation mechanisms and their respective contribution is still poorly understood. In particular, the presence of dark matter-deficient galaxies reported by observations \citep[e.g.,][]{vandokkum+18, vandokkum+19, mancerapina+19, guo+20, hammer+20} raises the need to search for different evolutionary scenarios other than the typical halo collapse model. It becomes crucial to consider the impact of the local environment and to study how external forces -- namely tidal interaction and ram pressure -- could impact existing galaxies and thereby invoke the formation of different kinds of galaxies, as observed objects are not always isolated. In that context, there have been several mechanisms proposed, such as tidal stripping of the dwarf's dark matter component by a massive companion or a cluster \refadd{\citep[e.g.,][]{ogiya+18, jing+19, niemiec+19, jackson+21, moreno+22}} and separation between the baryonic and non-interacting dark matter component during high-velocity collisions of several dwarf galaxies \citep[e.g.,][]{silk+19, shin+20, Lee+21, otaki_mori+23}.

Another pathway to produce dark matter-deficient objects discussed in the literature is the possibility of long-lived tidal dwarfs forming in the gaseous tails ejected by galaxy mergers \refadd{\citep[e.g.,][]{mirabel+92, bournaud_combes03, bournaud+04, bournaud_duc06,bournaud+08, kroupa+12}}. In such a scenario, structure formation would be triggered either by gas collapse due to Jeans instability or by local potential wells of ejected old stars. In both cases, this induces further gas accretion, triggering star formation and ultimately the birth of a new gravitationally bound and kinematically decoupled object \citep{duc12}. It is crucial to provide sufficiently large gas reservoirs to spark the formation of such star-forming pockets, suggesting that at least one of the merging galaxies needs to have a high gas mass fraction \citep{wetzstein+07}. Due to the inherent deficiency of dark matter at the tidal formation sites, such dwarf galaxies are expected to be dark matter-poor \citep{barnes_hernquist92}. Numerical studies of tidal dwarfs forming both in isolated mergers \citep{bournaud_duc06}, as well as in cosmological simulations \citep{ploeckinger+18, haslbauer+19} are able to produce dark matter-deficient tidal dwarf galaxies with a stellar population dominated by young stars. While isolated simulations predict that the produced tidal dwarfs generally will fall back into the merging galaxy pair, their development inside cosmological environments is still unclear, as the coarse temporal spacing between snapshots has not yet allowed a detailed investigation of the dwarfs' evolution. Although observations find tidal dwarfs with the aforementioned properties, for example those forming in the vicinity of \ngc{5291} \citep{duc_mirabel98, rakhi+23}, they also report cases of dwarf galaxies inside galaxy clusters with significant dark matter content or mixed stellar age component \citep[e.g.,][]{kaviraj+12, gannon+20, roman+21, gray+23}. However, it needs to be pointed out that unambiguously determining the tidal origin of dwarf galaxies in observations is difficult, especially once the optical bridges between progenitor and dwarf disappear. Afterwards, the only established observational fingerprint to identify isolated tidal dwarf galaxies are unusually high metallicities since their young stellar component forms from ejected, pre-enriched gas \citep{duc12}.

The presence of dwarf galaxies in galaxy clusters with particularly high metallicities for their luminosity \citep{rakos+00, duc+01, poggianti+01, iglesiasparamo+03} could signify a partly tidal origin, although different scenarios could also explain such peculiarities \citep{conselice+03b}. Simulating the formation and evolution of tidal dwarf galaxies is challenging as well, since the necessity to properly model gas fragmentation and succeeding star formation requires high resolution \citep{teyssier+10}. In case of an included environment, for example the hot gaseous atmosphere of a galaxy cluster, it is therefore necessary to resolve gas components of structures spanning many orders of magnitude in mass, driving up the numerical cost of the simulation. Nonetheless it becomes indispensable, as studies of single galaxies exposed to ram pressure already suggest a significant environmental impact by boosting the star formation activity, leading to so-called jellyfish galaxies \citep[e.g.,][]{kapferer+09, tonnesen+21, lee+22}.

The aforementioned system of \ngc{5291} in the western outskirts of cluster \abell{3574} is a convincing exemplary case to highlight the scientific importance of studying galaxy mergers inside cluster environments. It is an ongoing major merger between two galaxies -- the apparent early-type galaxy \ngc{5291} and its distorted companion referred to as \enquote{the \object{Seashell}} \citep{longmore+79}. The complex is surrounded by an extremely extended HI structure which is thought to be the gas-rich tidal arms stripped from the galaxies during the merger event. It features distinct star-forming knots towards the north and south of the system, which were first revealed by deep optical and spectroscopic studies \citep{gammelgaard+78,longmore+79} and later confirmed by observations in ultraviolet \citep[e.g.,][]{boqien+07,fensch+19,rakhi+23}. Such a constellation, combined with high metallicities and the absence of an old stellar population, suggests that these star-forming regions might be tidal dwarf galaxies \citep{malphrus+97,duc_mirabel98,bournaud+08}. As this merging system is situated within the outskirts of the galaxy cluster \abell{3574}, the cluster is suspected to have already impacted the ongoing merger, therefore clearly demonstrating the importance of studying how environmental effects are influencing the structures in order to be able to \refadd{reconstruct and predict} both their past and future behavior.

In this paper, we aim to answer the following questions: Is a cluster able to strip tidal dwarf galaxies from their progenitor? If so, how long would the stripped tidal dwarf galaxies survive while being exposed to the hostile conditions in a cluster? What general influence do these effects have on the probability of forming tidal dwarf galaxies? For this purpose, we perform hydrodynamical simulations of ongoing galaxy mergers in a cluster environment with \refadd{high mass resolution and unprecedented time resolution}, varying the infall direction between three different angles from a radial orbit to an initial infall angle of $45^\circ$. In \cref{sec:sim} we describe the simulation parameters and numerical implementation details. We provide a general proof of concept for environmentally supported tidal dwarf galaxy formation in \cref{sec:general_env_impact}. The detailed results are presented in \cref{sec:results}, where we analyze and discuss the emerging tidal dwarf population, as well as its temporal evolution. Furthermore, we use our results to provide an estimate of the tidal dwarf fraction among the total dwarf galaxy population inside galaxy clusters in \cref{sec:TDG_contribution}. Our conclusions are summarized in \cref{sec:sum_con}.

\section{Simulation}
\label{sec:sim}

The simulations \laned{are} performed using the Tree-SPH (smoothed particle hydrodynamics) code \textsc{Gadget-3}, which adopts the gravity force tree algorithm \citep{appel85, barnes_hut86} from its publically available predecessor \textsc{Gadget-2} \citep{springel05}, but employs an improved SPH implementation \cite[e.g.,][]{beck+16}. An SPH code decomposes the domain into \enquote{particles}, representative of the local mass distribution. The evolution of the system is then obtained by solving the hydrodynamic equations in Lagrangian form for each particle. Physical properties at each point can be retrieved by summing over the contributions of a specified number of neighboring particles, weighted by a radially symmetric smoothing kernel. In our simulations, we adopt smoothing over 64 neighboring particles with a Wendland C2 kernel \citep{wendland95,dehnen_aly12}.

The code includes the star formation and supernova feedback model by \cite{springel_hernquist03}, which treats the interstellar gas as a two-phase system with cold, star-forming gas embedded in an ambient hot medium. When the density of a gas particle exceeds a given threshold, it is converted into a stellar particle, representative of a stellar isochrone population according to an assumed initial mass function \citep[e.g.,][]{Salpeter1955}.
We incorporate radiative cooling of optically thin gas in ionization equilibrium with an ultraviolet background \citep{katz+96}. To follow the positions of the two galaxies, we place a massive tracer particle at each of the galaxy centers, representing a supermassive black hole. The initial conditions for the simulations \laned{are} created in three steps: initializing the late-type galaxies and combining them into a merger configuration (\cref{sec:galmerger_setup}), setting up the galaxy cluster (\cref{sec:cluster_setup})\laned{,} and combining the galaxy merger and the cluster into the final simulation setup (\cref{sec:final_setup}).

\begin{table*}[!ht]
    \centering
    \caption{\refadd{Model parameters and number of particles per type. The subscript in $N_i$ refers to dark matter (dm) and gas for both the galaxies and the cluster. The galaxies additionally contain stellar particles in the disk ($\ast, \rm disk$) and bulge ($\ast, \rm bulge$), as well as a massive, central tracer particle corresponding to a supermassive black hole (bh).}}
    \begin{tabular}{l r r r r r r r r}
        \hline\hline \\[-1em]
         Object & $M_{200}~ [10^{\refadd{11}}\,\Msun]$ & $R_{200}$ [kpc]& $V_{200}$ [km/s] & \refadd{$N_{\rm dm}$} & \refadd{$N_{\rm gas}$} & \refadd{$N_{\ast, \rm disk}$} & \refadd{$N_{\ast, \rm bulge}$} & \refadd{$N_{\rm bh}$}\\
        \hline \\[-0.9em]
        Galaxy a & $\refadd{9.5}$ & \refadd{203} & \refadd{142} & 265770 & 48060 & 12010 & 9520 & 1\\
        Galaxy b  & $\refadd{4.8}$ & \refadd{161} & \refadd{113} & 132890 & 24030 & 6010 & 4760 & 1\\
        Cluster & $1000.0$ & 957 & 670 & 50000000 & 50000000 & -- & -- & --\\
        \hline
    \end{tabular}
    \label{tab:virial_parameters}
\end{table*}

\subsection{Galaxy \laned{merger setup}}
\label{sec:galmerger_setup}
The individual disk galaxies were initialized using the method presented by \cite{springel+05}, which constructs a galaxy with the following components: a stellar disk with associated cold gas mass fraction $f_\text{gas}$, a stellar bulge, a dark matter halo, and a central black hole. The stellar and gaseous disks follow an exponential surface density profile given by
\begin{equation}
    \Sigma_i = \frac{M_i}{2\pi l_i} \exp \left(- \frac{r}{l_i} \right).
\end{equation}
The subscript $i$ represents the stellar ($\ast$) or gas component, whereas $l_i$ and $M_i$ denote the exponential scale length and total mass of the corresponding component. For the gas, we choose the scale length $l_\text{gas}$ to be twice as large as the scale length of the stellar disk since observations of late-type galaxies suggest that they have a larger radial extent of neutral hydrogen when compared to the stellar disk \citep[e.g.,][]{briggs+80, martin98}.

\refadd{An initialized galaxy is not in equilibrium at first and experiences an episode of intense star formation at the beginning of the simulation, effectively decreasing its gas mass fraction. In order to provide conditions favorable for tidal dwarf galaxy formation during the merger event \citep{wetzstein+07} and to reproduce the large hydrogen abundance in the merger complex consisting of \ngc{5291} and the \object{Seashell galaxy}, we set up gas-rich galaxies. Thus we choose the initial gas mass fraction in the disk (gas and stellar disk) to be $f_{\text{gas}}=0.8$, which corresponds to a gas to baryon (gas, stellar disk and bulge) mass fraction of $M_{\rm gas}/M_{\rm baryon}=0.5$. During the merger, when the tidal tails are most pronounced in our simulations, this fraction has dropped to $M_{\rm gas} / M_{\rm baryon}\approx0.3$. This corresponds to a gas mass of around $2\e{10}\Msun$ and coincides well with observational estimates of the total HI mass in the aforementioned merger complex \citep{malphrus+97}. We note that such gas-rich merger conditions are even more common at higher redshifts than at present-day, which enhances the likelihood for such an event to occur at higher redshifts.}

Both the bulge and dark matter components are modeled by the spherically symmetric Hernquist profile \citep{hernquist1990},
\begin{equation}
    \rho_k(r) = \frac{M_k}{2\pi} \frac{a_k}{r(r+a_k)^3},
\end{equation}
where the subscript $k$ denotes the bulge (b) or dark matter (dm) component, while $a_k$ and $M_k$ are the scale length and total mass of the respective halo. We parametrize the scale length of the bulge in terms of the disk scale length to $a_\text{b}=0.2 l_\ast$. In the case of the dark matter component, this density distribution \laned{is} selected since the total mass of a Hernquist profile converges and the associated analytical distribution functions allow an easier implementation than a NFW profile \citep{nfw1997}. To not lose touch with common descriptions of halos in cosmological simulations, $M_\text{dm}$ is set to be equal to the mass enclosed within $R_{200}$ in an NFW profile with matching central density, that is $\rho_{\text{NFW}} = \rho_{\text{dm}}$ for $r \ll R_{200}$, where $R_{200}$ is the radius at which the mean enclosed dark matter density is 200 times the critical density of the Universe. Using this condition, the scale length $a_\text{dm}$ is determined by the NFW concentration \mbox{parameter $c$} through \mbox{eq. 2} from \cite{springel+05}, which is set to $c=12$ for all \refadd{galaxy} halos.

In order to create a realistic setup we decided to perform preliminary tests, in which we searched for collisional configurations that reproduce the general characteristics of the aforementioned merger between \ngc{5291} and the \object{Seashell galaxy} inside the cluster \abell{3574}. In particular, we sought for arrangements leading to far-extended tidal features and a similar gas mass as in the observed counterpart. To this end, we finally chose a 2:1 prograde merger scenario, where the \refadd{more massive} galaxy has a total mass of \refadd{$M_{200, \rm gal}=9.5\e{11}\Msun$. The additional subscript (gal) indicates that we address the virial mass of a galaxy.} The total masses of the disk and bulge component as well as the central black hole mass are fixed fractions of the total halo mass \refadd{$M_{200, \rm gal}$}, being 0.041, 0.0065 and $1.186\e{-5}$, respectively. \refadd{The model parameters and the number of particles per type of the two initialized galaxies are listed in \cref{tab:virial_parameters}, where a and b refer to the larger and smaller galaxy of the 2:1 merger, respectively.} After initializing the galaxies, we used a routine provided by \cite{karademir+19} to create a merger configuration, which was later on placed in a cluster environment. The initial separation between the galaxies is \refadd{$\SI{140}{\kilo\parsec}$}, while their relative velocity is set to \refadd{144 km/s, which approximately corresponds to the virial velocity $V_{200, \rm gal}$ of the more massive galaxy}. We chose an impact angle of \refadd{$36^\circ$ between the two galaxies}, which corresponds to an impact parameter of \refadd{$\SI{82}{\kilo\parsec}$}.

\subsection{Cluster \laned{setup}}
\label{sec:cluster_setup}

To create initial conditions for the cluster environment, we \laned{used} the method described by \cite{donnert14}. In this model, the cluster consists of two components distributed spherically symmetrically, namely a dark matter halo and hot gas atmosphere, that is the intracluster medium (ICM). There is no need to include a stellar component of the cluster apart from the analyzed galaxy merger since the stellar component typically composes only a small fraction of the total cluster mass and does not interact with the infalling galaxy pair as it is collisionless. Following the same arguments as in \cref{sec:galmerger_setup}, the dark matter halo is set up by a Hernquist distribution, which closely follows an NFW-profile in the inner regions, but declines steeper at larger radii and thus results in a finite total mass. The ICM is modeled by the $\beta$-model \citep{cavaliere+1978}, which is motivated by observations \citep{croston+08}. Thus we have
\begin{equation}
    \rho_\text{ICM}(r) = \rho_0 \left( 1+\frac{r^2}{r_c^2} \right)^{-3\beta/2},
\end{equation}
where $\rho_0$ is the central gas density and $r_c$ the core radius of the gas distribution. Following the cluster simulations by \cite{mastropietro+08}, the exponent parameter is set to $\beta=2/3$. The baryon fraction inside $R_{200, \refadd{\rm cl}}$ of the cluster is $f_\text{bar}=0.14$. \refadd{The additional subscript (cl) in $R_{200, \refadd{\rm cl}}$ indicates that we address the virial radius of the cluster.}

For this parameter study, we chose the virial mass of the cluster to be \refadd{$M_{200, \rm cl}=10^{14}\,\Msun$}. We decided on this value since we wanted our simulated cluster to stay in touch with typical cluster masses such as that of the aforementioned cluster \abell{3574} \citep{saulder+16,oh+2018}, while also keeping the computation time at bay as the galaxy cluster at our desired resolution is quite expensive due to the large number of SPH particles. \refadd{The implemented parameters and number of particles per type for the initialized cluster are listed in \cref{tab:virial_parameters}.}

\subsection{Final \laned{simulation setup}}
\label{sec:final_setup}

\begin{table}[b]
    \centering
    \caption{Orbital parameters of our simulation sample.}
    \begin{tabular}{l c c}
        \hline\hline \\[-1em]
         Simulation &  Impact angle $\Theta$ & Impact parameter $d$ [Mpc]\\
        \hline \\[-0.9em]
        C0 & $0^\circ$  & 0 \\
        C25 & $25^\circ$ & 0.8\\
        C45 & $45^\circ$ & 1.3 \\
        \hline
    \end{tabular}
    \label{tab:configurations}
\end{table}

\begin{figure}[!b]
\centerline{\includegraphics[width=0.45\textwidth,trim={0 0 0 2cm},clip]{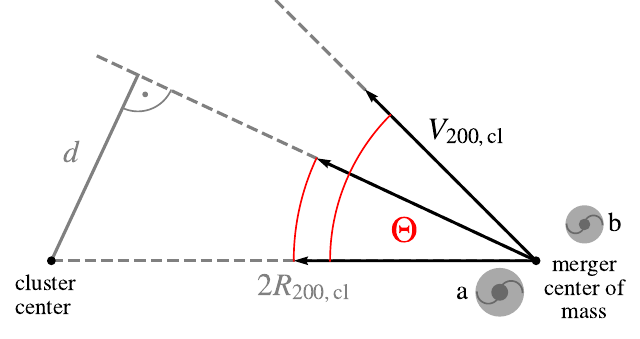}}
    \caption{Schematic setup of our simulation sample. We implement two disk galaxies at a distance of $2 R_{200,\refadd{\rm cl}}$ from the center of a galaxy cluster while varying the impact angle $\Theta$ with respect to the cluster, \refadd{resulting in an initial impact parameter $d$} (\cref{tab:configurations}). The initial velocity of the merging system with respect to the cluster is kept constant at $V_{200,\refadd{\rm cl}}$. $R_{200,\refadd{\rm cl}}$ and $V_{200,\refadd{\rm cl}}$ refer to the virial radius and velocity of the cluster.}
    \label{fig:cluster_config}
\end{figure}

\begin{figure*}[!t]
    \centerline{\includegraphics[width=0.355\textwidth]{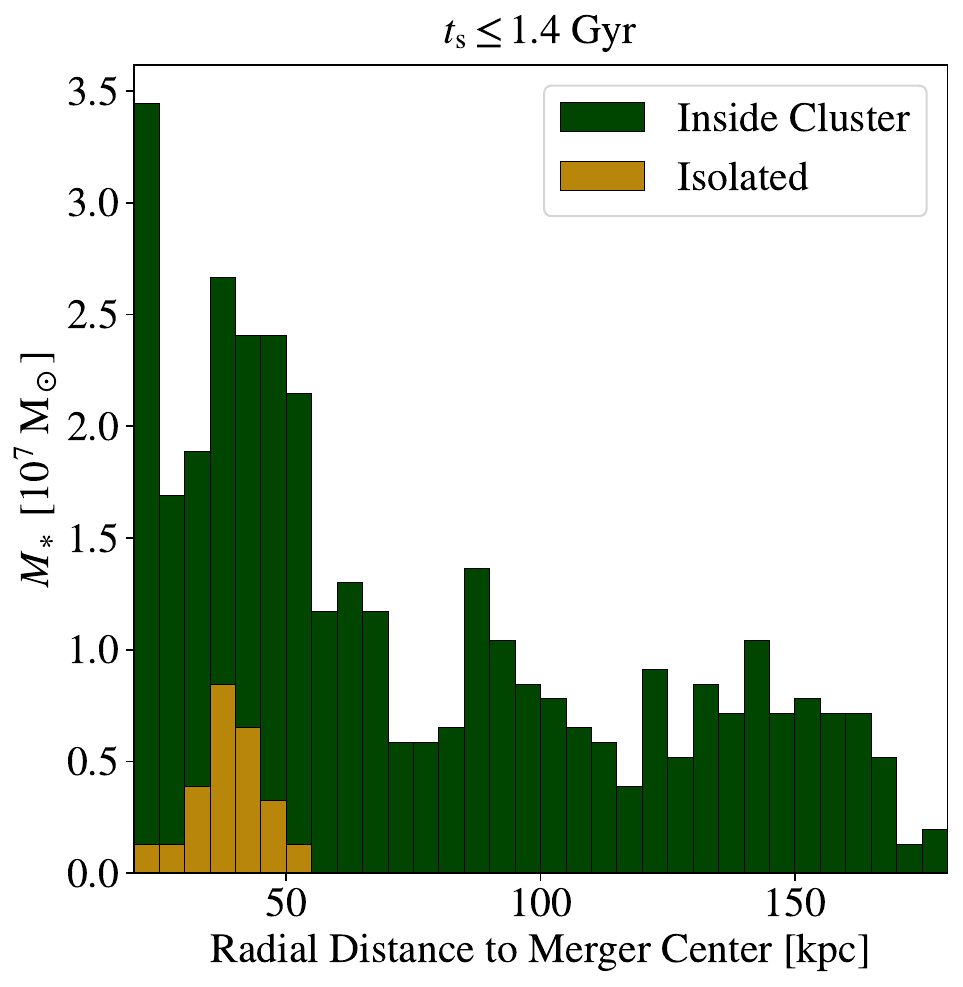}\includegraphics[width=0.62\textwidth,trim={0 -0.79cm 0 0}]{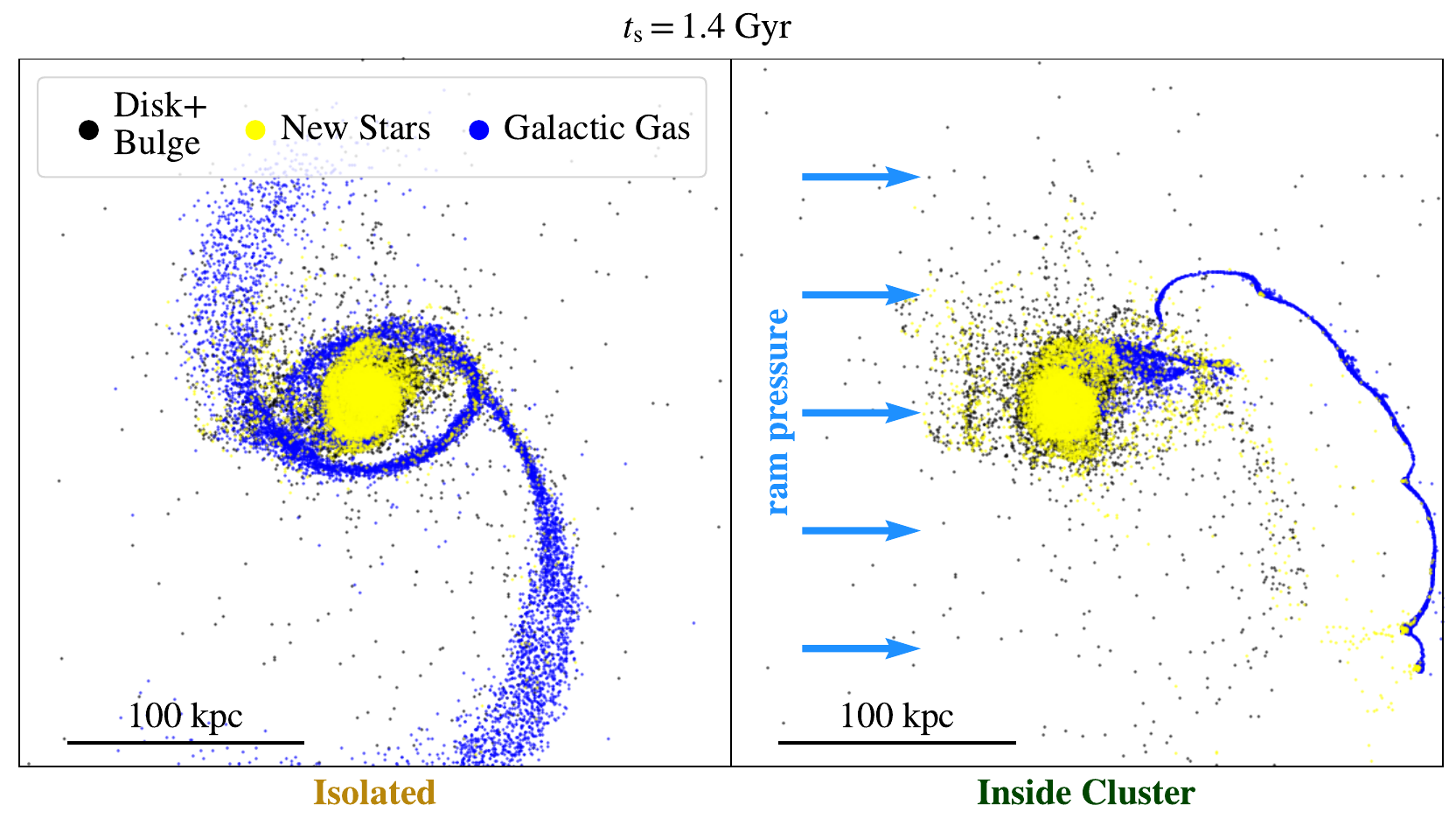}}
    \caption{\laned{Environmental impact on star formation activity and morphology of a galaxy merger.} \laned{Left}: \refadd{Stellar mass that formed within \SI{1.4}{\giga\year} after the beginning of the simulation in tidal tails as a function of distance from the progenitor's center of mass, at which the stellar particles were formed. The total formed stellar mass is summed within each radial bin}. The cases for galaxy mergers outside and within of a cluster environment are plotted in light brown and dark green, respectively. \laned{Center and right}: Morphology of the isolated galaxy merger (center panel) \SI{1.4}{\giga\year} after the beginning of the simulation, compared to the same merger but in a cluster environment (right panel) according to C0 in \cref{tab:configurations}. The old stellar component originally contained by the disk and bulge of the two merging galaxies is colored in black, while new stars that have formed since the beginning of the simulation are plotted in yellow. Blue markers show the position of gas particles initially contained in the disks of the galaxies, whereas the impact of the ICM is visualized by light blue arrows.}
    \label{fig:comparison_isolatedVSenv}
\end{figure*}

After initializing the participating objects, we needed to choose a realistic geometrical setup of the galaxy merger with respect to the cluster. Bound cluster members in the inner regions of a galaxy cluster typically exhibit high relative velocities, causing mergers to be improbable events. But while galaxies are still in the filamentary inflow region towards the cluster, their relative velocity is expected to be much lower than for their infallen counterparts, allowing them to undergo collisions. Hence, mergers that are found inside galaxy clusters are thought to have fallen in together as bound structures. Therefore, we chose the starting distance of the galaxy merger to be at $2R_{200,\refadd{\rm cl}}\approx \SI{1.9}{\mega\parsec}$, while their initial velocity with respect to the cluster was taken to be its virial velocity $V_{200,\refadd{\rm cl}}\approx 670$ km/s. To investigate the orbital impact on the emerging satellite dwarf population, we \laned{focus} on three setups in this paper, which differ in the impact angle of the merger with respect to the cluster. Meanwhile, the merger configuration of the two disk galaxies and their orientation towards the cluster center remain the same. The geometrical setup is illustrated in \cref{fig:cluster_config}, while the different orbital parameters \refadd{(impact angle $\Theta$ and corresponding impact parameter $d$)} of our simulation sample are listed in \cref{tab:configurations}. \refadd{We name the simulations C$x$, where $x$ is denoting the implemented impact angle in the respective configuration.}

We \laned{choose} the same resolution for all gas particles in the simulation volume, \refadd{namely} for those initially associated to the galaxies, as well as for those contained by the cluster. Even though this leads to computationally expensive simulations because of a large amount of gas particles contributed by the cluster, we \laned{make} this decision to properly resolve the gas interaction between the cluster and the galaxies without contaminating the results with numerical artifacts due to different resolutions between the SPH particles. The stellar resolution is the same as for the gas, so for all baryonic particles the particle masses are set to \mbox{$m_\text{bar}=6.5\e{5}\,\Msun$}. The dark matter resolution is set to \mbox{$m_\text{dm}=3.2\e6\,\Msun$}. We \laned{choose} the softening lengths for the baryonic and dark matter component to be $\epsilon_\text{bar}= \refadd{\SI{9}{\parsec}}$ and $\epsilon_\text{dm}=\refadd{\SI{39}{\parsec}}$ , respectively. For this, we applied the values used in existing galaxy merger simulations \citep{johansson+09a,johansson+09b,remus+13} and scaled them according to our resolution using $\epsilon \propto m^{1/3}$, where $\epsilon$ is the softening length and $m$ the particle mass. Hence, this approach is based on the assumption that the interparticle distance upon initialization scales with $m^{1/3}$ as well. The total simulation volume is a cube with box length $2 \times 3.75 R_{200, \refadd{\rm cl}}= \SI{7.2}{\mega\parsec}$\refadd{, while the time step between consecutive snapshots is \SI{14}{\mega\year}}.

\section{Environmental \laned{impact}}
\label{sec:general_env_impact}

\begin{figure*}[!b]
    \centerline{\includegraphics[height=0.4\textwidth,trim={-8cm 0 0 0cm}]{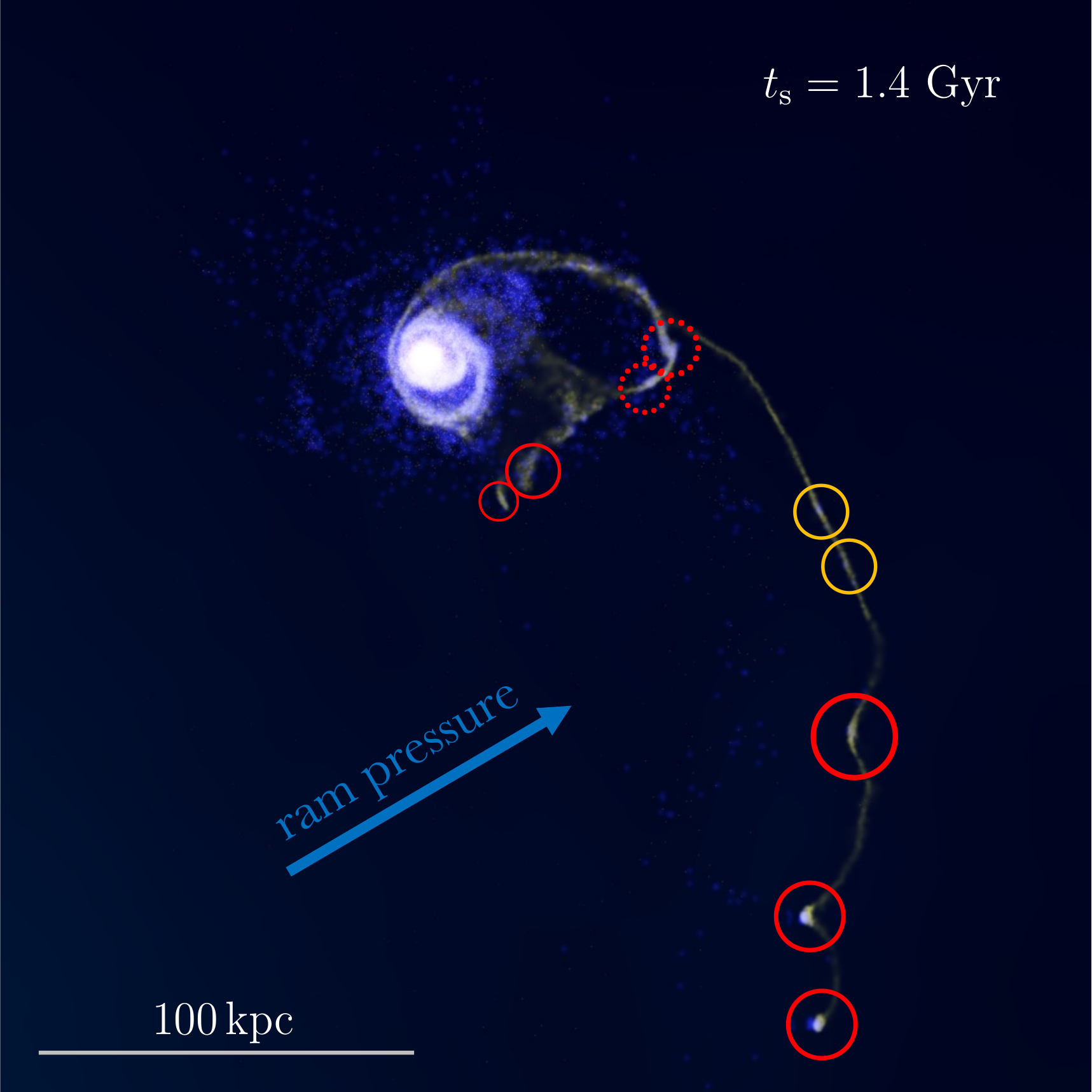}\hspace{1mm}\includegraphics[height=0.41\textwidth,trim={0 3mm 0 0},clip]{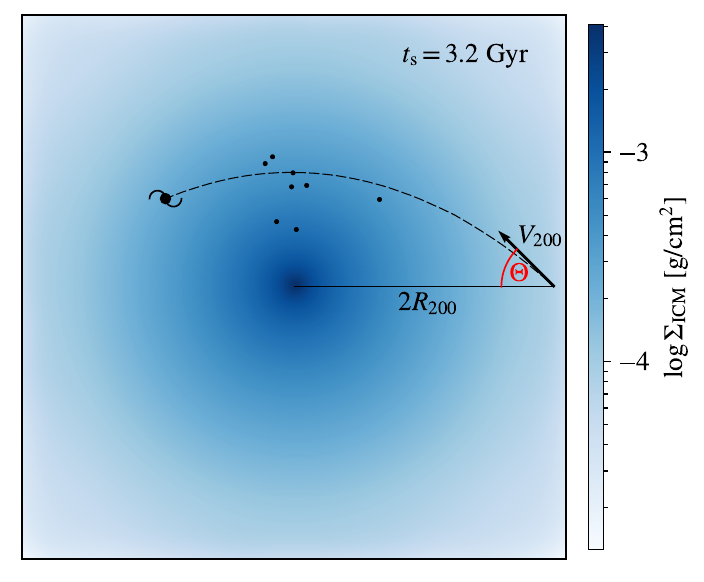}}
    \caption{\laned{Emergence of tidal dwarf galaxies in galaxy clusters.} \laned{Left}: Simulated galaxy merger in a cluster (C45) \SI{1.4}{\giga\year} after the beginning of the simulation. Gas and young stars are rendered in yellow and light blue, respectively, using the raytracing visualization software Splotch \citep{dolag+08}. Circles highlight star forming pockets, which either fall back into the merger (dashed), or managed to escape the local gravitational potential (solid), while \refadd{orange} circles mark two dwarfs which coincidentally merge after the stripping event. \laned{The associated movie, showing the merger and the stripping process of tidal dwarf galaxies, is available at \href{https://www.youtube.com/watch?v=EQyEK1qQAhU}{this URL}.} \laned{Right}: Spatial distribution of stripped tidal dwarfs (black dots) formed by the galaxy merger \SI{3.2}{\giga\year} after the beginning of the simulation. The dashed line traces the trajectory of the galaxy pair, while the blue background indicates the integrated surface density of the ICM in the plotted plane.}
    \label{fig:simsnap}
\end{figure*}

Before presenting a detailed analysis of the dwarf galaxies obtained from the simulation, we provide a proof of concept for environmentally supported tidal dwarf formation. Placing a galaxy merger into a cluster with hot ambient medium drastically enhances the star formation activity across the tidal features, as becomes apparent from the histogram in \cref{fig:comparison_isolatedVSenv}. \refadd{It compares the star formation activity occurring inside the tidal tails of an isolated galaxy merger (light brown) to the case of a galaxy merger inside a galaxy cluster (dark green), which was set up according to C0 in \cref{tab:configurations}. While $t_{\rm s} = 0$ represents the beginning of the simulation, the histogram shows the stellar mass forming inside the tidal tails until $t_{\rm s} = \SI{1.4}{\giga\year}$ against the radial distance from the merger's center, at which the stellar particle is formed. The total stellar mass is summed within each radial bin.} We define the stellar tidal tails as the star-forming regions that are sufficiently far away from the closest black \mbox{hole ($>20\, \text{kpc}$)} in order to exclude the active merger center. The elevated density inside the tidal tails exposed to the ICM leads to significantly higher star formation rates compared to the isolated merger. Additionally, we can see how the environment enables star formation at much \refadd{larger} distances to the merging progenitors.

The momentary morphology of the merger \refadd{at $t_{\rm s} = \SI{1.4}{\giga\year}$} is shown next to the histogram in \cref{fig:comparison_isolatedVSenv} for the isolated \refadd{(center panel)} and environmental \refadd{(right panel)} case. Both mergers are set up exactly the same -- the only difference is the cluster environment of the merger in the right \refadd{panel}. We do not plot the gas of the ICM here in order to highlight the arising morphological differences between the two cases, particularly for the galactic gas ejected by the merging galaxies shown in \refadd{blue}. Light blue arrows indicate the direction in which ram pressure is  impacting the galaxy merger in the environmental case. The old stellar population, which initially belonged to the disk and bulge components of the galaxies, is plotted in \refadd{black}, while new stars that have formed after the beginning of the simulation are indicated in yellow. Moving through the ICM, ram pressure becomes a crucial influence, leading to a significantly different morphological evolution. This continuous push causes the tidal tails to become less spread-out and denser than in the isolated case. While the upper tidal feature disperses entirely, the lower one has a smaller relative velocity with respect to the ICM due to the merger's sense of rotation (anti-clockwise in the plotting plane). Hence, it experiences a less rapid head wind and manages to harbor its gas, while also undergoing substantial compression. This process eventually triggers the genesis of several star-forming pockets across the whole tidal complex, representing new born, gas dominated tidal dwarf galaxies. Assisted by ram pressure, they later are able to escape the progenitor's gravitational potential.

\section{Analysis of \laned{stripped tidal dwarf galaxies}}
\label{sec:results}

\begin{figure*}[b]
\centerline{\includegraphics[width=0.95\textwidth,trim={0 0 0 0},clip]{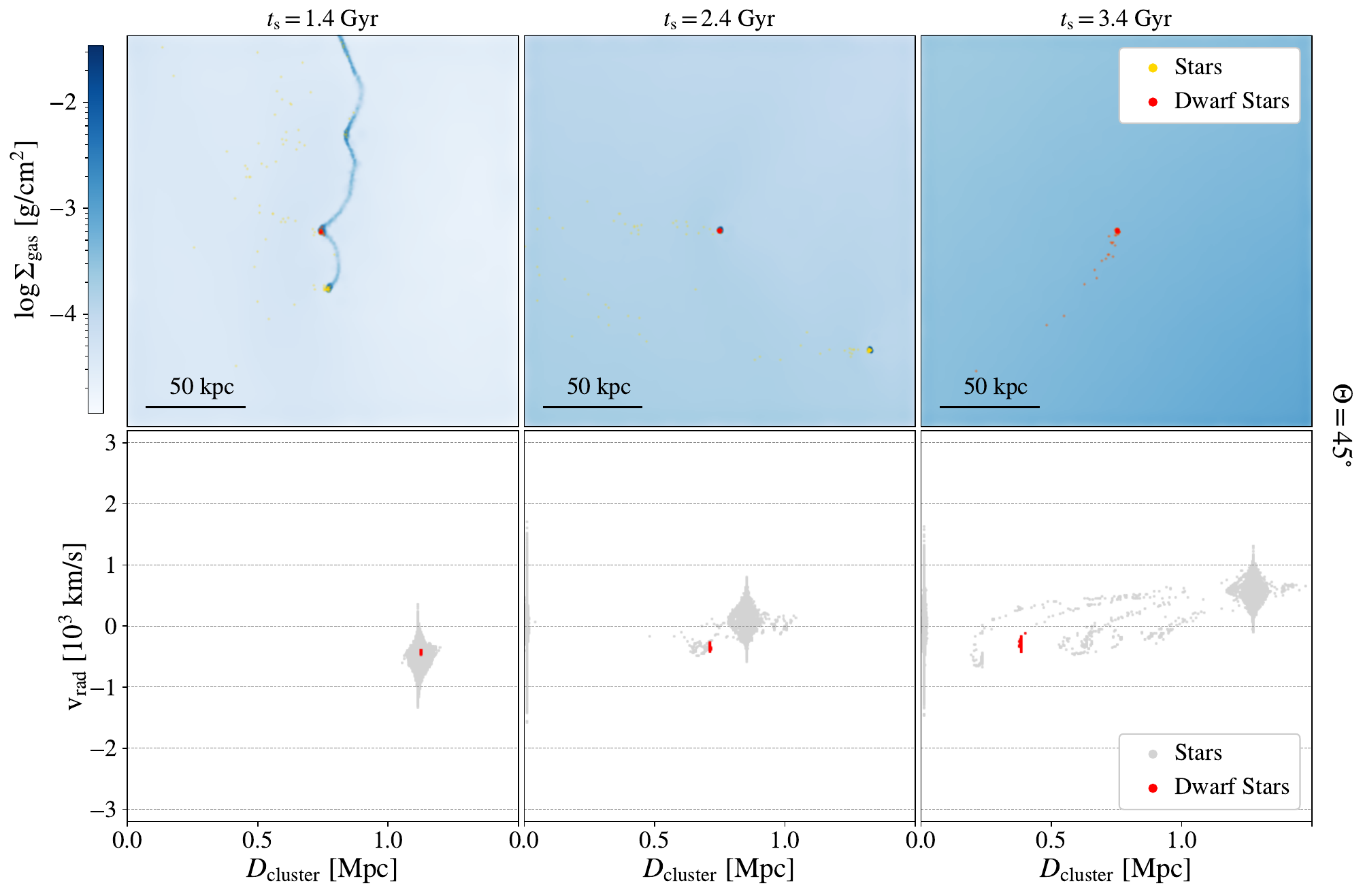}}
\caption{Real space (upper row) and phase space (lower row) distributions of a dwarf forming on the elliptical merger orbit C45 for three different snapshots in its evolution with \SI{1}{\giga\year} time steps between consecutive panels, as indicated above the columns. The stellar content of the dwarf, identified at the moment of the middle panel, is traced in red. The integrated surface density of the gas is visualized in blue, while stars not belonging to the dwarf are plotted in yellow and gray in real and phase space, respectively.}
\label{fig:phasespace_45}
\end{figure*}

\begin{figure*}[htp]
    \centering
    \begin{subfigure}{0.7\textwidth}
    \includegraphics[width=1\textwidth]{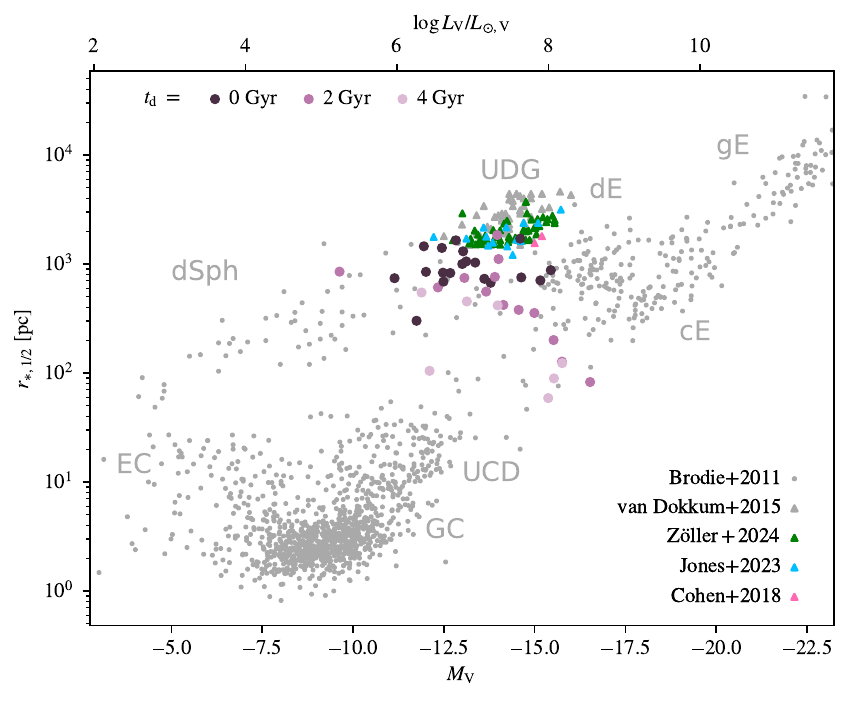}
    \includegraphics[width=1\linewidth]{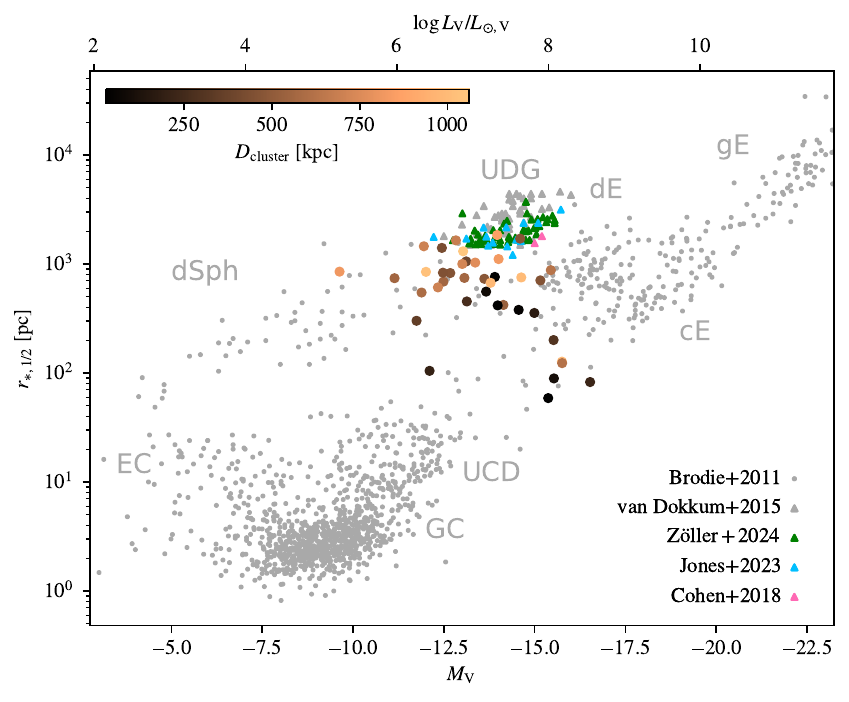}
    \end{subfigure}
\caption{\refadd{Half-light radii vs.} visual band luminosities for various object types. Gray circles show stellar systems observed in the local Universe including giant, compact, and dwarf ellipticals (gE, cE, and dE), dwarf spheroidals (dSph), ultra-compact dwarfs (UCD) as well as globular and extended clusters (GC and EC) compiled by \cite{brodie+11}. \refadd{Triangles indicate ultra-diffuse galaxies (UDG) observed by various authors: UDGs inside the Coma cluster by \cite{vanDokkum+15} and \cite{zoeller+23} in gray and green, respectively; gas-rich field UDGs by \cite{jones+23} in light blue; and finally \ngc{1052}-\object{DF2} and \object{DF4} observed by \cite{cohen+18} in light red.} \laned{Top}: Purple colors indicate stripped tidal dwarf galaxies in all three simulation samples (C0, C25, C45), while the shade encodes the time span that has passed since the beginning of our dwarf tracing. \laned{Bottom}: The same objects as in the upper panel, but colored according to their radial distance from the cluster center.}
    \label{fig:sages}
\end{figure*}

Tidal dwarf galaxies are formed and stripped from the merging galaxies in all three orbit configurations presented in \cref{sec:final_setup}, which vary in the merger's impact angle with respect to the cluster. During this process significant compression of tidal gas occurs, as seen in the left panel of \cref{fig:simsnap}, which shows a snapshot of configuration C45 at $t_{\rm s}\approx \SI{1.4}{\giga\year}$. \refadd{As before, $t_{\rm s} = 0$ is representing the beginning of the simulation.} Such behavior is induced by the merger event and enhanced by ram pressure through the intracluster gas. This process gives birth to pockets with high star formation rates, highlighted by circles in \cref{fig:simsnap}. Although there are tidal dwarfs that eventually fall back into their progenitor (dashed circles), the majority of them are able to escape the local potential of the merging galaxy pair and start to evolve independently (solid circles). Curiously, we also find two stripped tidal dwarfs in this simulation configuration that form a rotationally supported binary system, which later collapses into a single dwarf galaxy (orange circles). In the radial infall scenario C0, all dwarf structures are quickly destroyed after about \mbox{1--2\,Gyr} by tidal forces upon passage through the cluster center. For the elliptical orbits C25 and C45, however, \refadd{some} dwarf galaxies manage to survive much longer for up to \mbox{$\sim$\SI{4}{\giga\year}}, while spiraling \refadd{towards} the cluster center. We \laned{stop} the dwarf tracing at that point because most objects then had either reached the cluster center, or had small cluster-centric distances at that time. 

\subsection{Emerging \laned{dwarf population}}
\label{sec:emerging_dwarf_pop}

\begin{figure*}[b]
\centerline{\includegraphics[width=0.33\textwidth]{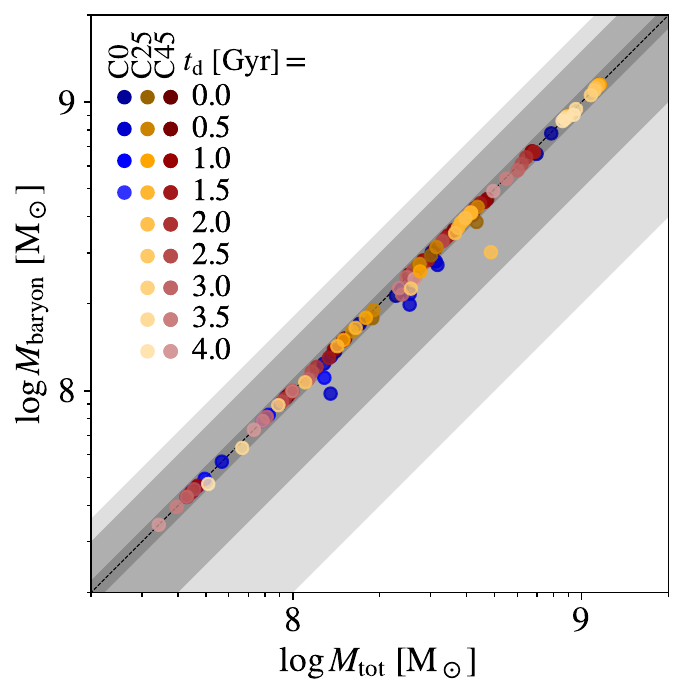}\includegraphics[width=0.33\textwidth]{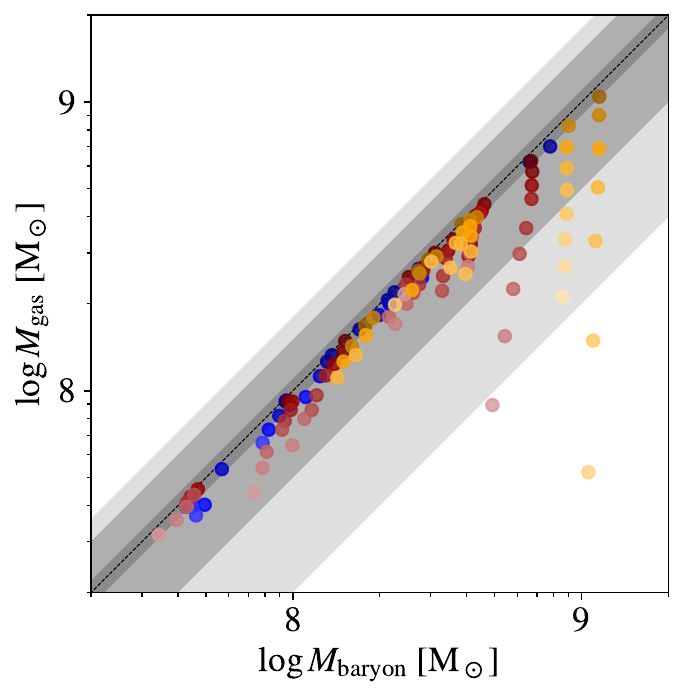}\includegraphics[width=0.33\textwidth]{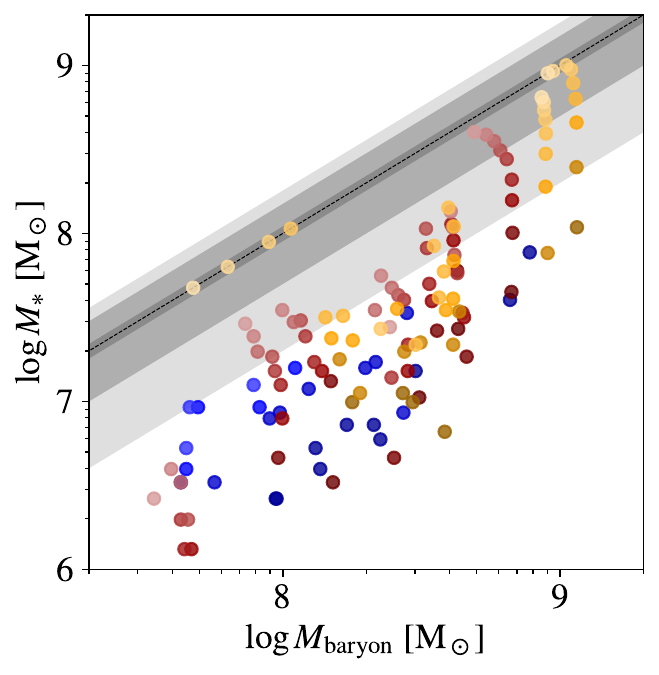}}
\caption{Baryon mass $M_\text{baryon}$ vs.\ total mass $M_\text{tot}$ (left), gas mass $M_\text{gas}$ vs.\ baryon mass $M_\text{baryon}$ (middle), and stellar mass $M_\ast$ vs.\ baryon mass $M_\text{baryon}$ (right) of stripped tidal dwarf galaxies. The blue, orange and red colors represent the C0, C25 and C45 orbit scenarios, respectively. Values at the beginning of the tracing are shown in dark colors, while their evolution within \SI{4}{\giga\year} with time steps of \SI{0.5}{\giga\year} is indicated by consecutively lighter shades. The dashed black line indicates the 1:1 relationship, while the gray regions represent a 10\%, 50\%, and 80\% decline from it.}
    \label{fig:mgas_vs_mtot}
\end{figure*}

By visually inspecting the simulation in real and phase space, we identify the position and extent of dwarf galaxies that are disassociated from the galaxy merger remnant. \refadd{To find possible dwarf candidates, we extract structures with at least \laned{five} stellar particles and trace them forward in time. This corresponds to a stellar mass of about $3\e{6} \Msun$. Although this is a low threshold, we point out that only one dwarf in our sample displayed such a small mass and that all other objects started out more massive. If they do not disperse and exhibit stellar growth, we recognize them as dwarf galaxies in our sample. We choose to search for these dwarfs in the simulation by visual inspection, rather than using algorithms such as \textsc{SubFind} \citep{springel+01,dolag+09} in order to find spatially coherent structures as one would in observations. The major advantage of this approach is that we are able to find tidal dwarf candidates early in their evolution while they are still gas-dominated, since \textsc{SubFind} tends to recognize subhalos with significant stellar content more reliably in case of small particle numbers and large tidal fields. For completeness, however, we have applied \textsc{SubFind} to our simulations after our analysis. We confirm that only a minority ($\approx 20\%$) of the tidal dwarfs found by visual inspection and that survive for at least \SI{1}{\giga\year} are classified as formally not gravitational bound by \textsc{SubFind}. However, these dwarf galaxies still remain spatially coherent due to their stars' similar orbit inside the cluster.}

We begin to analyze the dwarf galaxies shortly after the first stellar clumps started forming in the tidal tail of the merger, as they first need to get stripped before they can be identified as isolated dwarfs. Throughout this work, $t_{\rm d}=0$ represents the beginning of the dwarf tracing, which is at $t_{\rm s}\approx \SI{1.7}{\giga\year}$ after the beginning of the simulation. \refadd{A dwarf is traced across the snapshots and we record its properties every \SI{0.5}{\giga\year}.} We find \laned{three, seven, and eight} dwarfs present at $t_{\rm d}>\SI{1}{\giga\year}$ for the infall angles $\Theta=0^\circ, 25^\circ$, and $45^\circ$, respectively. In the two latter scenarios, \laned{three} and \laned{four} dwarf galaxies are still present at $t_{\rm d}=\SI{4}{\giga\year}$.

\Cref{fig:phasespace_45} displays an example of a dwarf formed by a galaxy merger on the $\Theta = 45^\circ$ orbit in the cluster. The top and bottom rows in \cref{fig:phasespace_45} show the distribution in real and phase space, respectively, while the columns are snapshots at three different times. \refadd{In red we plot stellar particles that were identified as dwarf galaxy at the moment of the middle panel. The left and right hand side panels, on the other hand, show the distribution of these stars \SI{1}{\giga\year} before and after this moment, respectively. Therefore, there are less red markers in the left hand side panels, since a part of the identified stars has not formed yet.} It is clearly visible how the stellar content of the traced dwarf is seeded in the high-density regions of the tidal tail (top left panel) and is continuously fueled by active star formation after getting stripped. The dwarfs do not keep their full stellar body, though, as they develop a leading tail with time (top right panel). Comparing the position in phase space between the dwarf and the merger remnant, which is the largest symmetric structure, it becomes apparent that the tidal dwarf gets dynamically decoupled (bottom right panel).

The population around $\refadd{D_{\rm cluster}}=0$ in the lower panels of \cref{fig:phasespace_45} corresponds to stars forming at the cluster center, since by construction the cluster initially does not have a central galaxy, and hence, the gas starts to cool and ignite star formation there. This population would be part of the brightest cluster galaxy (BCG) if such a galaxy would have been included in the simulation. However, as we consider a dwarf to be disrupted when it passes the cluster center, we do not incorporate a BCG in order to save computation time. A similar analysis, but for a dwarf from the radial infall scenario $\Theta = 0^\circ$, is supplemented in the Appendix in \cref{fig:phasespace_0}.

\begin{figure*}[!b]
    \centerline{\includegraphics[height=0.45\textwidth,trim={0 0 0 0},clip]{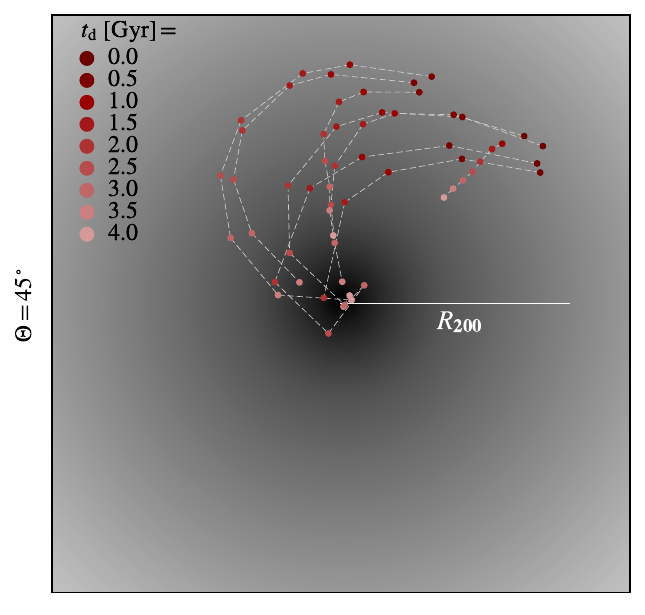}\includegraphics[height=0.45\textwidth,trim={0 0 0 0},clip]{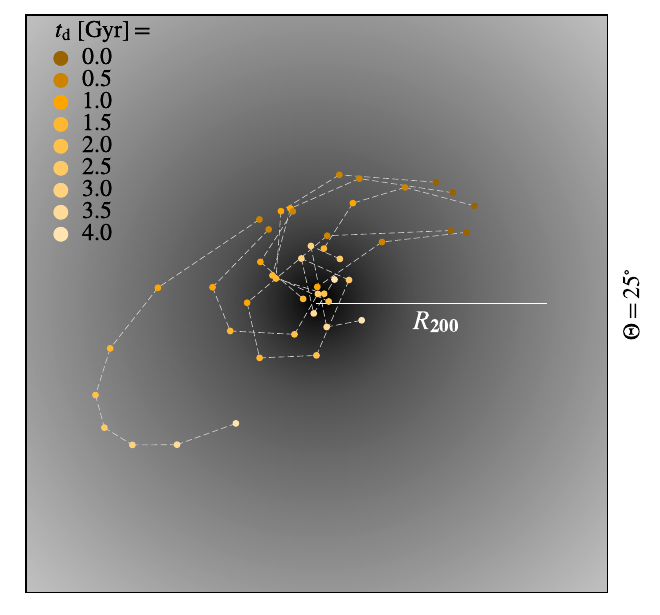}}
\caption{Tracks of stripped tidal dwarf galaxies in a cluster over a time period of \SI{4}{\giga\year}. The left and right panels show the two elliptical orbit cases C45 and C25, respectively. Consecutively lighter shades indicate progressively later times.}
    \label{fig:spatial_distribution}
\end{figure*}

To understand what kind of objects are forming, we compare the simulated dwarfs with a broad range of galaxies observed in the nearby Universe. \Cref{fig:sages} shows how different types of galaxies populate distinct regions in the stellar size-luminosity plane. The vast majority of those observed objects, spanning from giant ellipticals to stellar clusters, was compiled by \cite{brodie+11}, indicated by gray circles. In addition, we also include data of ultra-diffuse galaxies (UDGs) for comparison, indicated by triangles. \refadd{UDGs in the Coma cluster by \cite{vanDokkum+15} and \cite{zoeller+23} are plotted in gray and green, respectively, while light blue markers represent gas-rich field UDGs by \cite{jones+23}. Additionally, we show the two UDGs \ngc{1052}-\object{DF4} and \object{DF2} observed by \cite{cohen+18} in light red. These two galaxies} \laned{are} speculated to have formed through a high-velocity collision of gas-rich galaxies \citep{Lee+21, vanDokkum+22}. In the upper panel, we plot the stripped tidal dwarf galaxies from all three simulations (C0, C25, and C45) in purple for three different times in their evolution, which are each \SI{2}{\giga\year} apart and distinguished by consecutively lighter shades. For comparison, we also provide the orbit-encoded distribution of our sample in the Appendix (\cref{fig:sages_app}). \refadd{While the observed stellar radii are projected values, the stellar half-mass radii $r_{\ast,1/2}$ of the simulated dwarfs are computed from the three-dimensional stellar distribution, in order to avoid introducing biases by choosing a specific plane. Since \cref{fig:sages} covers a large range of stellar radii, this would in any case lead to only marginal changes of the simulated dwarfs' positions inside the figure.} The luminosity of the simulated dwarf galaxies was estimated by applying age-dependent mass-to-light ratios provided by \cite{sextl+23} for each stellar particle in the simulation. \refadd{This means that older stellar particles \laned{were} attributed with lower luminosities than younger particles, which encapsulates the dimming of a stellar isochrone with time since the most massive and luminous stars are also the first to decease.}

The simulated tidal dwarf galaxies mostly populate a range from dwarf ellipticals (dE) to dwarf spheroidals (dSph), with some even reaching \refadd{values} of ultra-compact dwarfs (UCD). Interestingly, we find a small overlap between observed UDGs and our simulated dwarf galaxies at the large-size end. This overlap is mostly observed at earlier stages during the first \SI{2}{\giga\year} of their evolution, since the dwarfs tend to decrease in stellar size during their infall into the cluster, and hence, become more compact with time, moving towards the bottom in \cref{fig:sages}. Simultaneously, their luminosity increases due to ram-pressure boosted star formation, resulting in an overall trend of dwarfs moving \refadd{diagonally to the bottom right} in the size-luminosity relation. This becomes even clearer by the lower panel of \cref{fig:sages}, where the dwarf galaxies are colored according to their distance from the cluster center. It can be seen immediately that dwarfs closer to the cluster center tend to be more compact, while dwarfs farther away are larger and more diffuse.

Such behavior raises the intriguing question as to whether low-luminosity galaxies might undergo a transformation from the diffuse stellar population towards compact objects. In their study of galaxy properties in the \object{Hydra I cluster}, \cite{LaMarca_1+22} \laned{find} a weak correlation of declining stellar radii for decreasing distance to the cluster center. Additionally, \cite{jannsens+19} \laned{report} a clear anticorrelation between the spatial distribution of UDGs and UCDs in the \laned{Hubble} Frontier Fields galaxy clusters, which are much more massive than \object{Hydra I}. There, UCDs are more concentrated in the center, while UDGs are more often found in the outer regions within the virial radius. Concurrently, clusters with the lowest intracluster light (ICL) also \laned{show} the least amount of UCDs. Since tidal disruption is commonly assumed to be the leading cause for the emergence of ICL \citep[e.g.,][]{burke+12}, \cite{jannsens+19} \laned{hypothesize} that UCDs in the central regions of a galaxy cluster might form through tidal stripping of infalling UDGs. Given that they also \laned{find} a non-uniform spatial distribution of dwarf galaxies around cluster cores of non-relaxed clusters, this would further suggest that these objects could be associated with infalling substructures.

These observations are in excellent agreement with the trends found in our simulations. While starting out as diffuse objects at early times, the simulated dwarf galaxies begin to experience environmental stripping. Meanwhile, the gas becomes more condensed, which enhances the star formation inside the dwarf core. This results in increasingly compact stellar distributions with time. Since they are moving towards the cluster, this transformation is correlated with their cluster-centric distance. Hence, our simulation results confirm that such an evolutionary pathway of dwarf galaxies is indeed possible.

\subsection{Mass \laned{evolution}}
\label{sec:mass_evolution}
Since they originate from the gas tail of the galaxy merger, the stripped tidal dwarfs are generally dark matter-deficient and exhibit high star formation rates due to their gas dominance. This is shown in \cref{fig:mgas_vs_mtot}, which depicts the dwarfs' mass evolution for their different components. The three tested orbit scenarios of the galaxy merger falling into a cluster with an angle of $\Theta = 0^\circ, 25^\circ$, and $45^\circ$ are visualized by blue, orange and red markers, respectively. Consecutively lighter shades represent progressing time between $t_{\rm d}=\text{0--4\,Gyr}$ with time steps of \mbox{$\sim$\SI{0.5}{\giga\year}}. The dashed line shows the 1:1 relationship. In order to guide the eye, we indicate a 10\%, 50\%, and 80\% decline from it with gray backgrounds. Hence, if a dwarf lies, for example, on the lower edge of the lightest gray region, it contains 20\% of its mass in the component on the y-axis with respect to the component on the x-axis.

The leftmost figure displays the baryon mass $M_\text{baryon}$ (stars and gas) versus total mass $M_\text{tot}$ (stars, gas, and dark matter). All dwarfs lie close to the 1:1 relationship, indicating that they are completely baryon-dominated. The reason is that these objects originate from the outskirts of the galaxy merger's potential, where little dark matter is available to inherit. Since they are constantly exposed to ram pressure, these gaseous dwarf galaxies experience high star formation rates, continuously converting their gas reservoir into stars. This behavior can be deduced from the middle panel, showing the gas mass $M_\text{gas}$ versus baryon mass $M_\text{baryon}$ of the tidal dwarfs. Interestingly, the most massive dwarfs also experience the highest star formation rates, visible by the downwards evolution within this figure. \refadd{Their large sizes facilitate the formation of cold star-forming clouds embedded in the dwarf galaxy, which is surrounded by the hot ICM, while also stabilizing the dwarf against tidal shear in the cluster.} The slight migration towards smaller baryon masses indicates mass loss to the environment as well. The rightmost figure further demonstrates how the dwarf galaxies increase their stellar mass $M_\ast$ with time, eventually even resulting in complete domination of the stellar component for the most massive objects. Thus, dwarf galaxies originating from the tidal tails of a merger can \refadd{evolve into} gas-free, stellar dominated \refadd{objects}, if the survival time in the cluster potential is large enough.

\subsection{Evolution in \laned{detail}}

Dwarf galaxies forming in the radial infall scenario (C0) quickly reach the cluster center due to a lack of initial angular momentum, whereas objects forming on elliptical merger orbits (C25, C45) naturally survive for a longer period of time. This behavior is visualized in \cref{fig:spatial_distribution}, which shows the time dependent position of the traced dwarf sample with respect to the cluster center for the infall angles $\Theta=45^\circ$ (left panel) and $\Theta=25^\circ$ (right panel). The different shades of red (C45) and orange (C25) indicate the motion of the dwarfs  with time. It becomes apparent that most dwarf galaxies spiral inwards while continuously losing momentum due to friction between the dominating gas component of the galaxies and the intracluster gas, indicated by the gray background. The orbit of the merging galaxy pair dictates the initial angular momentum of the tidal dwarf galaxies, naturally determining their future orbit trend inside the cluster.

Curiously, there is an outlier in terms of dynamical behavior in C45, which adapts a radial orbit and moves much slower than the rest. This is the smallest dwarf galaxy in  that sample, which barely surpasses the resolution threshold, having a final gas and stellar mass of $M_\text{gas} = 4\e{7} \,\Msun$ and $M_\ast = 10^6~ \Msun$, respectively. The comparably small mass causes the dwarf to lose its angular momentum due to friction on much shorter timescales, while ram pressure also simultaneously slows down the radial infall of the gaseous dwarf galaxy. A massive object, on the other hand, is less impacted by environment, such as the dwarf in C25 tracing a hook-like orbit, resulting in the largest final cluster-centric distance within the sample. Hence, this dwarf galaxy could even survive significantly longer than the simulation time of \SI{4}{\giga\year}.

\begin{figure}
    \centerline{\includegraphics[width=0.5\textwidth]{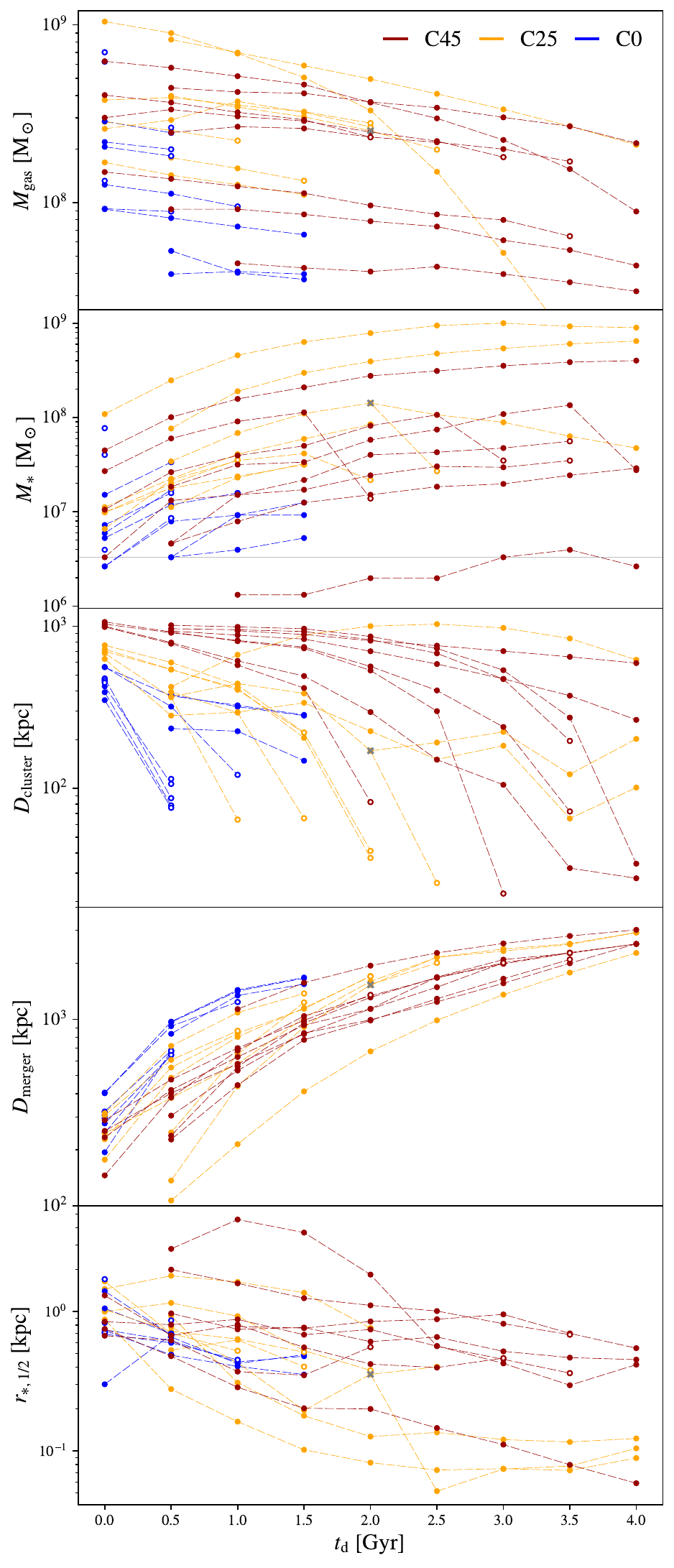}}
\caption{Evolution of stripped tidal dwarf galaxies forming due to a galaxy merger on three different orbits in a cluster (C0, C25, and C45). From top to bottom, we show gas mass $M_\text{gas}$, stellar mass $M_\ast$, cluster-centric distance $D_\text{cluster}$, distance $D_\text{merger}$ to the galaxy merger remnant \refadd{and stellar half-mass radius $r_{\ast,1/2}$ of the dwarfs}. Open circles indicate that the dwarf is destroyed afterwards. The gray horizontal line in the second panel represents a resolution threshold of five stellar particles.}
    \label{fig:against_t}
\end{figure}

In \cref{fig:against_t} we present the evolution of the gas mass $M_\text{gas}$, the stellar mass $M_\ast$, the distance from the center of mass of the cluster $D_\text{cluster}$, the distance to the merger remnant $D_\text{merger}$ \refadd{and the stellar half-mass radius $r_{\ast,1/2}$} of the three simulated scenarios with cluster infall angles $\Theta=0^\circ$ (blue), $\Theta=25^\circ$ (orange), and $\Theta=45^\circ$ (red). Open circles indicate that the dwarf galaxy was not present anymore at the subsequent tracing time. Dwarfs forming in the wake of a galaxy merger moving radially towards the cluster (C0) do not survive longer than $t_{\rm d} =\SI{1.5}{\giga\year}$ because they reach the cluster center at that point. Comparing the gas mass with the stellar mass evolution, we see a clear trend of decreasing gas masses, while the stellar masses rise due to active star formation. The abrupt decline in stellar mass of some dwarfs is explained by close passage, and hence, tidal stripping by the cluster center, after which many are destroyed, even if they do not pass the center directly. Other disrupted dwarfs do not show the decline in stellar mass but rather just vanish. These are dwarfs with lower initial masses, and hence, are dismantled much quicker, which is not resolved with the coarse time steps of \mbox{$\sim$\SI{0.5}{\giga\year}}. At $t_{\rm d} = \SI{4}{\giga\year}$, we still find \laned{three} and \laned{four} surviving dwarf galaxies with tidal origin for the two elliptical orbit scenarios C25 and C45, respectively. At this point, they have already reached large distances to their progenitor on the order of the cluster's virial radius (third panel in \cref{fig:against_t}). These showcases demonstrate that isolated dwarf galaxies in clusters can be of tidal origin and be present for a significant fraction of the Hubble time.

Each of the two elliptical cases C25 and C45 displays a noticeable outlier beginning its lifetime at the high gas mass end, which then converts gas into stars at exceptionally higher rates compared to the other dwarfs in its respective sample. \refadd{As mentioned before in \cref{sec:mass_evolution}, their large sizes stabilize them against tidal torques. The comparably deep potential wells of these dwarf galaxies then lead to dense accumulations of gas, which are further compressed by ram pressure -- ultimately leading to intense star formation.} We also mention another peculiar object -- a dwarf in the C25 configuration -- whose stellar and gas components are sheared into two separate objects at \mbox{$t_{\rm d}=\SI{2}{\giga\year}$}. At that point, it has reached a stellar mass of $1.4\e{8}\Msun$, marked by a gray cross in \cref{fig:against_t}. One of the descendants is entirely made of stars with no gas component and continues to be present in the cluster, whereas the other inherits the available gas reservoir but evaporates after \mbox{$t_{\rm d}=\SI{2.5}{\giga\year}$}. The former descendant explains a peculiarity in the baryon mass $M_{\rm baryon}$ vs.\ stellar mass $M_\ast$ relation in \cref{fig:mgas_vs_mtot}, namely the completely stellar-dominated object from C25 (orange) around $M_{\rm baryon}\approx 10^8~ \Msun$, traced for four time steps.

Although their stellar mass keeps rising, almost all dwarf \refadd{galaxies} exhibit a decreasing stellar radius over time \refadd{(see lowest panel in \cref{fig:against_t})}, as already mentioned in \cref{sec:emerging_dwarf_pop}. \refadd{Additionally, dwarfs from the C25 sample reach smaller sizes than their counterparts in the C45 sample.} We argue that this is mainly the consequence of \refadd{stripping and} ram pressure. Weakly bound gas in the outskirts of a dwarf is continuously stripped away, leaving only gas in the center, which can form stars. \refadd{This effect is enhanced at smaller cluster-centric distances, which explains the systematically smaller sizes of the C25 sample. Its dwarfs reach the cluster center faster compared to dwarfs from the C45 simulation (c.f. \cref{fig:spatial_distribution}) due to the smaller impact angle of the initialized galaxy merger.} Additionally, a trail of stars, which we find in many of the traced dwarfs (cf.\ upper right panel of \cref{fig:phasespace_45}), points towards a lag of the dwarf's gas with respect to its stellar component. Ram pressure slows down the gaseous dwarfs while stars do not experience this influence. Hence, we find stars from the dwarf's outskirts \enquote{hurrying ahead}, since they are less bound than stars at the potential center. As a result, the dwarfs only keep stars at their very center, which results in small stellar radii.

\section{Contribution of \laned{tidal dwarf galaxies to total dwarf population in galaxy clusters}}
\label{sec:TDG_contribution}
While early studies estimated that all dwarf galaxies could be of tidal origin \citep{okazaki_taniguchi00}, follow-up surveys deduced much smaller fractions: 16\% in the observational sample by \cite{sweet+14}, 10\% by \cite{bournaud_duc06}, where the authors simulated isolated galaxy major mergers, and even as low as 6\% by \cite{kaviraj+12}, who performed a statistical observational investigation of tidal dwarf galaxies in the local Universe. The latter study employed the result of \cite{bournaud_duc06} that in conditions favorable for tidal dwarf formation -- being wet, prograde major mergers with mass ratios from 1:1 to 4:1, and inclinations between the two orbital planes $\leq$40$^\circ$ -- only 1--2 massive ($M_\text{tot}>10^8\,\Msun$) tidal dwarfs per merging galaxy \laned{survived} for at least \SI{1}{\giga\year}. However, this conclusion is only valid in the case of a field environment since the simulation sample consisted of isolated galaxy mergers. The main reason for the low number of tidal dwarfs per merger in their study \laned{was} that the only structures able to survive for a long period of time \laned{were} those that \laned{had} high initial masses and \laned{formed} far enough away from the progenitor galaxies, such that they \laned{were} not disrupted by the merger's tidal field. In our work, we \laned{show} that the number of massive tidal dwarf galaxies formed by a major merger can be significantly higher when environmental interaction with \refadd{an} ICM is taken into account. Since the aim \laned{is} a proof of concept using three exemplary orbit configurations, our dwarf production rate is not a robust statistical prediction. Still, to demonstrate the potential increase of the true tidal dwarf galaxy fraction in the local Universe, we repeat the estimation performed in section 6 of \cite{kaviraj+12}, but use the dwarf formation rate from our simulations \refadd{in the following}.

The observational sample studied by \cite{kaviraj+12} \laned{suggested} that only \mbox{$\sim$18\%} of wet major mergers \laned{produced} tidal dwarf galaxies. In conditions that allow long-lived tidal dwarfs -- that is no radial infall and thus no rapid tidal disruption by the cluster -- we produce \mbox{$\sim$6.5} stable (present for $>\SI{1}{\giga\year}$), massive ($M_\text{tot}>10^8\Msun$) tidal dwarf galaxies in the configurations C45 (6 dwarfs) and C25 (7 dwarfs). Hence, we estimate that $0.18 \times 6.5=1.17$ \laned{tidal dwarfs form per wet, major merger event}. By integrating an empirically motivated major merger rate, \cite{consolice07} concluded that each massive galaxy experiences \mbox{$\sim$4} major mergers in its lifetime. Since early interactions before $z=1$ are expected to be dominated by gas-rich galaxies \citep[e.g.,][]{kaviraj+09} and at most one major merger usually happens after that time \citep[e.g.,][]{conselice+03a, lin+04, jogee+09} this indicates that, statistically, a massive galaxy experiences \mbox{$\sim$3} wet, major mergers. Considering earlier works \citep{bournaud10}, which found that \mbox{$\sim$50\%} of massive tidal dwarf galaxies (i.e., $M_\text{tot}>10^8\,\Msun$) \laned{survived} \refadd{for} a significant fraction of the Hubble time \refadd{(present for several \laned{gigayears})} we arrive at $3 \times 0.5 \times 1.17=1.76$ \laned{long-lived tidal dwarfs per massive galaxy}. The observed galaxy mass function of the Coma cluster \citep{secker_harris96} suggests a number ratio between dwarf galaxies with $M_{\rm tot} > 10^8\,\Msun$ to massive galaxies of \mbox{$\sim$5.8}. Using this value, we therefore conclude that the ratio of tidal galaxies among dwarfs in clusters could be as high as \mbox{$\sim$30\% ($= 1.76/5.8$)}.

We stress that several oversimplifying assumptions \laned{enter} this estimate. As pointed out by \cite{kaviraj+12}, this approach assumes a constant tidal dwarf production rate over cosmic time, even though mergers at higher redshift are expected to be more gas-rich, amplifying the probability of forming tidal dwarf galaxies. Moreover, we \laned{assess} the probability (namely \mbox{$\sim$50\%}) of the dwarfs' long-time survival by applying the statistical result from the isolated merger simulations of \cite{bournaud_duc06}, in which the reason for short lifetimes was tidal disruption by the progenitor. Although the limiting factor in our sample is the hostile cluster environment, this is only of minor concern since the survival fraction in our simulations is in a similar range. Furthermore, the orbit of the galaxy merger in the cluster \laned{is} not taken into account. It determines both the survival rate of tidal dwarfs being disrupted in the cluster center, as well as the impact angle between the tidal tails and the direction of ram pressure, which can lead to a different number of gaseous dwarf galaxies per merger being able to escape the local gravitational potential. The size of the galaxy cluster could also play a significant role since we expect a more massive cluster to impose stronger ram pressure caused by the increased ICM density, as well as by the possibly increased infall velocities of the merging galaxies relative to the ICM.

\refadd{Compared to studies on tidal dwarf galaxies in cosmological simulations \citep[e.g.,][]{ploeckinger+18}, we find significantly more dwarfs per merger, directly leading to an increased fraction of dwarf galaxies that could be of tidal origin. It is important to note, however, that these studies were performed on relatively small cosmological boxes, because otherwise the resolution would not suffice to study such small objects as tidal dwarf galaxies. Since the box size determines the scale of the largest possible collapsed node, this means that galaxy clusters in these boxes will be relatively small, if present at all. Additionally, a cluster will form later than an equal mass counterpart in a larger simulation box, having less time to relax and to develop a hot gas atmosphere. These aspects could lead to lower ram pressure acting on satellite galaxies, which we identify as the crucial driver for the formation and stripping of tidal dwarf galaxies in cluster environments. Therefore, statistical analyses of tidal dwarf galaxies in large cosmological boxes with simultaneously high resolutions are necessary in order to determine their true contribution towards the total dwarf galaxy population. With these things in mind,} our result demonstrates that the fraction of tidal dwarf galaxies is most probably higher than currently adopted in the literature. \refadd{Our deduced value of $\sim$30\% is not the final prediction, but rather a benchmark on how large the fraction could be when taking into account more realistic environments produced by large cosmological simulations.}

\section{Summary and Conclusion}
\label{sec:sum_con}
Inspired by observations of numerous newly formed tidal dwarf galaxies originating from the merger event between the galaxy \ngc{5291} and \enquote{the \object{Seashell}} inside the galaxy cluster \abell{3574}, we addressed the following questions: Can galaxy cluster environments enhance the formation of tidal dwarf galaxies? \refadd{Can such tidal dwarf galaxies be stripped from their merging hosts by a cluster environment and thus survive longer than those formed by an isolated galaxy merger -- thereby explaining observed populations of dark matter-free dwarf galaxies seen in many galaxy clusters?} To that end, we performed and analyzed a sample of three idealized simulations in extremely high resolution as a proof of concept, consisting of a galaxy major merger in a cluster environment with varying impact angle towards the cluster center.

We demonstrate that the environment is indeed able to trigger star formation in the tidal tails of the merger at large distances and subsequently strip these newly formed objects from the merger potential. Decoupled from their cradle, the tidal dwarf galaxies begin to evolve independently with increasing distances from the merger remnant up to \laned{megaparsec} range. Since they originate from the gas tail of the merger, these stripped tidal dwarfs are generally dark matter-deficient. Hence, the baryonic component dominates their total mass. Exposed to ram pressure, these objects experience high star formation rates, while simultaneously decreasing their stellar half-mass radii, resulting in diverse dwarf galaxy types. Consequently, we \laned{observe} a general trend of diffuse dwarf galaxies becoming more compact as their distance to the cluster center decreases. The gas and stellar mass of the found dwarfs range between $M_\text{gas} \approx 10^{7-9}\,\Msun$ and $M_{\ast} \approx 10^{6-9}\,\Msun$, respectively, while the stellar half-mass radii typically lie between \mbox{$r_{\ast1/2}\approx 10^{2-3}\, \rm pc$}. We traced the evolution of the stripped tidal dwarfs over a time period of \mbox{$\sim$\SI{4}{\giga\year}} after the merger event, in which the dwarf galaxies spiral towards the cluster center with varying timescales. Although a fraction of these objects is eventually destroyed by the tidal field, we still find distinguishable dwarf galaxies after \mbox{$\sim$\SI{4}{\giga\year}} with cluster-centric distances on the order of \SI{10}{\kilo\parsec} to a few \SI{100}{\kilo\parsec}.

Since the initial conditions of our simulations were orientated by the merger between \ngc{5291} and the \object{Seashell galaxy}, our results predict that the star-forming knots observed in the merger's tidal tails could transform into independent satellite dwarf galaxies of the cluster in the future. Compared to the isolated merger simulations by \cite{bournaud_duc06}, we find a significantly higher tidal dwarf production rate for mergers inside galaxy clusters due to environmentally supported formation. Based on our results, we repeated a similar computation as done by \cite{kaviraj+12} and predict that the fraction of dwarf galaxies with tidal origin could be on the order of \mbox{$\sim$30\%}. The \refadd{tidal dwarf production rate per merger} may be even higher in more massive clusters than the one simulated in this work since such circumstances might cause higher ram pressure acting on the galaxies' gas. This prospect, as well as the impact of varying subgrid physics implementations, will be subject of future investigations but are beyond the scope of this proof-of-concept study.

To conclude, we find that tidal dwarf galaxies formed during mergers of galaxies within a galaxy cluster environment have a higher probability to be stripped from the merger potential, while their stellar masses are enhanced due to environmental effects. Such tidal dwarf galaxies are dark matter-deficient and can distribute over the whole cluster. Most importantly, they appear in a full variety of observed dwarf galaxy types, from diffuse disky dwarfs to ultra-diffuse galaxies but also ultra-compact galaxies. Therefore, this channel of dwarf galaxy production inside a galaxy cluster is a possible mode to form dark matter-deficient dwarf galaxies of all kind, and can explain several unsolved issues currently discussed with regard to dwarf galaxies inside galaxy clusters. Disentangling the details that lead to the different dwarf galaxy types in \refadd{a} galaxy cluster environment will thus be an important endeavor \refadd{addressed in future works}.

\begin{acknowledgements}
We thank Eva Sextl and Rolf Kudritzki for kindly providing us with age-dependent mass-to-light ratios that we used in the compilation of \cref{fig:sages}. \refadd{We thank the anonymous referee for their helpful comments and suggestions, which greatly contributed to the quality of this work.} AI and KD acknowledge support by the COMPLEX project from the European Research Council (ERC) under the European Union’s Horizon 2020 research and innovation program grant agreement ERC-2019-AdG 882679. KD also acknowledges support by the Deutsche Forschungsgemeinschaft (DFG, German Research Foundation) under Germany’s Excellence Strategy - EXC-2094 - 390783311. LMV acknowledges support by the German Academic Scholarship Foundation (Studienstiftung des deutschen Volkes) and the Marianne-Plehn-Program of the Elite Network of Bavaria. The following software was used for this work: Julia \citep{bezanson+17:julia}, Splotch \citep{dolag+08}.
\end{acknowledgements}

\bibliographystyle{style/aa_url}
\bibliography{bib}

\begin{appendix}

\section{Phase space evolution for \texorpdfstring{$\Theta=0^\circ$}{Theta=0deg} (C0)}
\label{sec:phasespace}

\begin{figure}
\centering
\includegraphics[width=0.49\textwidth,trim={0 0 0 0},clip]{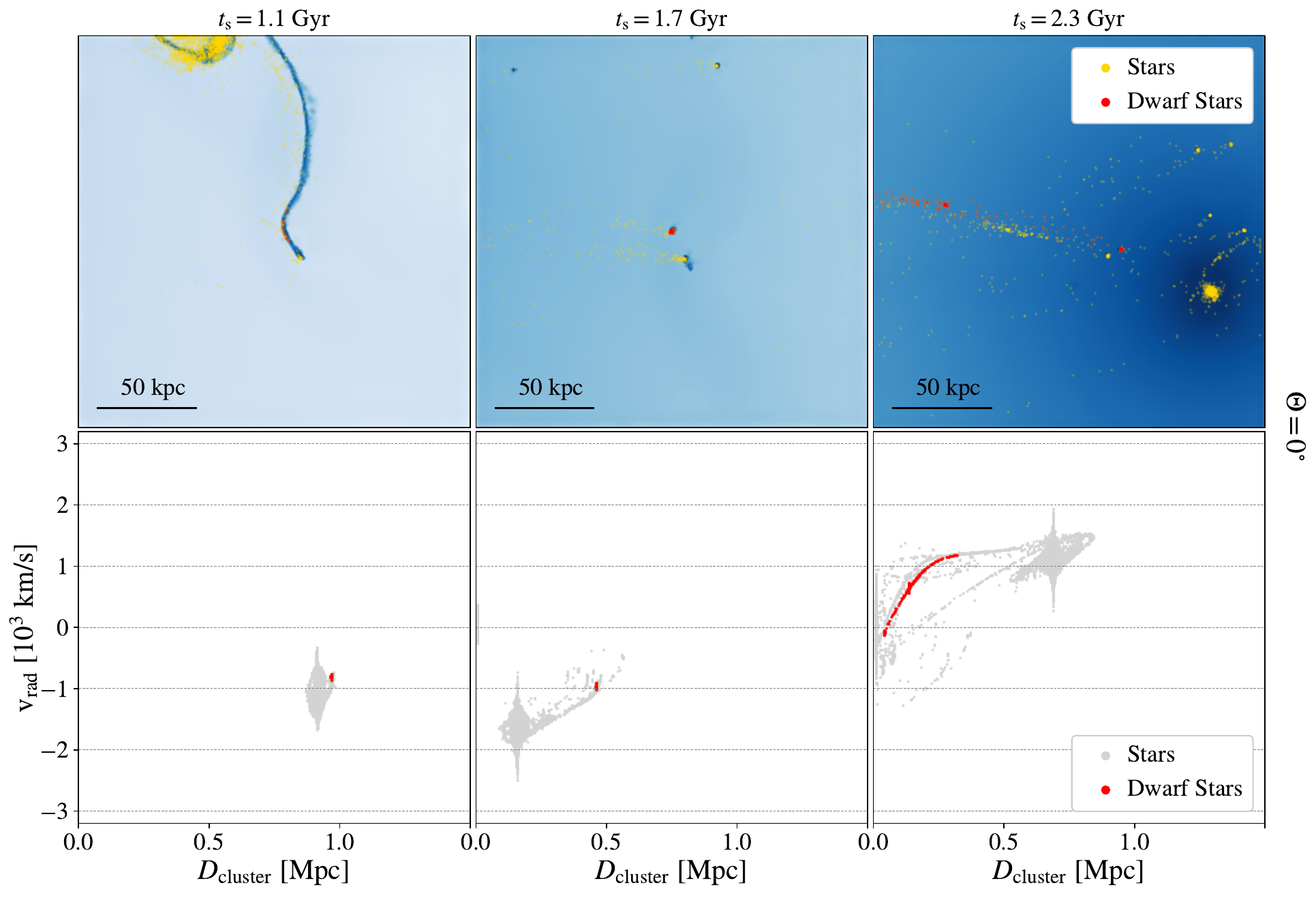}
\caption{Real space and phase space distribution in the top and bottom row, respectively, of a dwarf forming in the radial infall merger configuration C0 for three different snapshots in its evolution with \SI{0.6}{\giga\year} between consecutive panels. The stellar content of the dwarf, identified at the moment of the middle panel, is traced in red. The integrated surface density of the gas is visualized in blue, while stars are plotted in yellow and gray in real and phase space, respectively. This figure is complementary to \cref{fig:phasespace_45}.}
\label{fig:phasespace_0}
\end{figure}

We include in \cref{fig:phasespace_0} the evolution in real and phase space for a dwarf from the infall scenario C0 for comparison, as an addition to the elliptical orbit case C45 presented in \cref{fig:phasespace_45}. Objects in the radial scenario naturally evolve on shorter timescales than the dwarfs on elliptical orbits, which is why the time separation in \cref{fig:phasespace_0} is about half as long compared to \cref{fig:phasespace_45} in order to properly visualize the environmental impact on the dwarf galaxy. Having almost no angular momentum with respect to the cluster, the progenitor as well as its tidal dwarf galaxies are destined to fall into this massive central potential well, visible as the dark blue spot in the latest time panel in \cref{fig:phasespace_0}. As a result, all structures are \enquote{stretched} out both in real and phase space by the tidal field, which even rips the traced dwarf apart into two separate remains.

\section{Emerging Dwarf Population Distinguished by Orbit and Cluster-Centric Distance}

\begin{figure}
\centering
\includegraphics[width=0.49\textwidth]{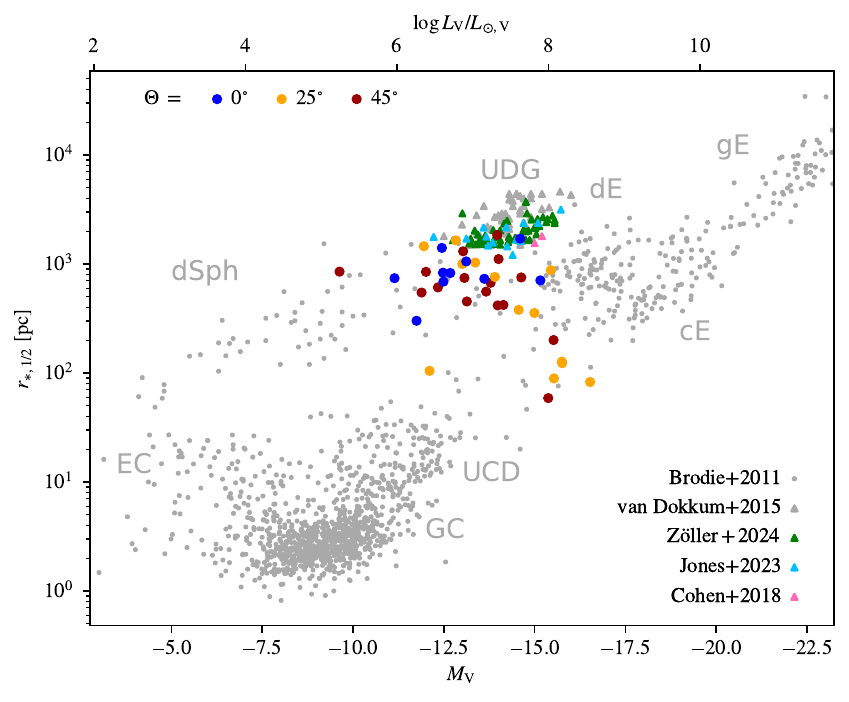}
\caption{\refadd{Simulated dwarf galaxy distribution at the same three moments in time as in \cref{fig:sages}, but with different color coding. Blue, orange, and red circles represent the C0, C25, and C45 orbit scenarios, respectively. For comparison, we include observational data by various authors listed in the legend.}}
\label{fig:sages_app}
\end{figure}

For completeness we provide the dwarf galaxy distribution in the stellar size-luminosity plane for the three moments in time as presented in \cref{fig:sages}, but apply different color coding in \cref{fig:sages_app}. Here, we distinguish \refadd{the stripped tidal dwarf galaxies by the orbit scenario C0, C25 and C45 in blue, orange and red, respectively}. The dwarfs from the radial infall scenario C0 mostly populate the large size end. This is because they are all destroyed by tidal torques after reaching the cluster center within $\sim$\SI{1.5}{\giga\year} and thus do not follow the general trend of decreasing stellar sizes with time visible for the other two simulated dwarf galaxy populations, as described in the main text.

\end{appendix}

\end{document}